\documentclass[twocolumn,pre,superscriptaddress,amsmath,amssymb,longbibliography]{revtex4-1}

\usepackage{graphicx}
\usepackage{dcolumn}
\usepackage{bm}
\usepackage{float}
\usepackage{color}

\begin{document}


\title{Three-Dimensional Construction of Hyperuniform, Nonhyperuniform and Antihyperuniform Random Media via Spectral Density Functions and Their Transport Properties}

\author{Wenlong Shi}
\affiliation{Materials Science and Engineering, Arizona State
University, Tempe, AZ 85287}
\author{Yang Jiao}
\email[correspondence sent to: ]{yang.jiao.2@asu.edu}
\affiliation{Materials Science and Engineering, Arizona State
University, Tempe, AZ 85287} \affiliation{Department of Physics,
Arizona State University, Tempe, AZ 85287}
\author{Salvatore Torquato}
\email[correspondence sent to: ]{torquato@electron.princeton.edu}
\affiliation{
Department of Chemistry, Princeton University, Princeton, New Jersey 08544, USA}
\affiliation{
Department of Physics,  Princeton University, Princeton, New Jersey 08544, USA}
\affiliation{
Princeton Institute  of Materials, Princeton University, Princeton, New Jersey 08544, USA}
\affiliation{
Program in Applied and Computational Mathematics, Princeton University, Princeton, New Jersey 08544, USA}



\date{\today}

\begin{abstract}

Rigorous theories connecting physical properties of a heterogeneous material to its microstructure offer a promising avenue to guide the computational material design and optimization. The spectral density function ${\tilde \chi}_{_V}({k})$, which can be obtained experimentally from scattering techniques, enables accurate determination of various transport and wave propagation characteristics, including the time-dependent diffusion spreadability ${\cal S}(t)$ and effective dynamic dielectric constant $\epsilon_e$ for electromagnetic wave propagation. Moreover, ${\tilde \chi}_{_V}({k})$ has been employed to derive sharp bounds on the fluid permeability $k$. Given the importance of ${\tilde \chi}_{_V}({k})$, we present here an efficient Fourier-space based computational framework to construct {\it three-dimensional} (3D) statistically isotropic two-phase heterogeneous materials corresponding to targeted spectral density functions. In particular, we employ a variety of analytical ${\tilde \chi}_{_V}({k})$ functions that satisfy all known necessary conditions to construct disordered stealthy hyperuniform, standard hyperuniform, nonhyperuniform, and antihyperuniform two-phase heterogeneous material systems at varying phase volume fractions $\phi$. We show that by tuning the correlations in the system across length scales via the targeted functions, one can generate a rich spectrum of distinct structures within each of the above classes of materials. Importantly, we present the first realization of antihyperuniform two-phase heterogeneous materials in 3D, which are characterized by a power-law autocovariance function $\chi_{_V}(r)$ and contain clusters of dramatically different sizes and morphologies. We also determine the diffusion spreadability ${\cal S}(t)$ and estimate the fluid permeability $k$ associated with all of the constructed materials directly from the corresponding ${\tilde \chi}_{_V}({k})$ functions. Although it is well established that the long-time asymptotic scaling behavior of ${\cal S}(t)$ only depends on the functional form of ${\tilde \chi}_{_V}({k})$, with the stealthy hyperuniform and antihyperuniform media respectively achieving the most and least efficient transport, we find that varying the length-scale parameter within each class of ${\tilde \chi}_{_V}({k})$ functions can also lead to orders of magnitude variation of ${\cal S}(t)$ at intermediate and long time scales. Moreover, we find that increasing solid volume fraction $\phi_1$ and correlation length $a$ in the constructed media generally leads to a decrease in the dimensionless fluid permeability $k/a^2$, while the antihyperuniform media possess the largest $k/a^2$ among the four classes of materials with the same $\phi_1$ and $a$. These results indicate the feasibility of employing parameterized ${\tilde \chi}_{_V}({k})$ for designing composites with targeted transport properties.

\end{abstract}



\maketitle


\section{Introduction}

Heterogeneous materials, such as composites, multi-phase alloys, porous media, granular matters, and biological gels and tissues, to name but a few, are ubiquitous in nature and synthetic forms \cite{torquato2002random, Sa03a}. Such media are of great importance in a wide spectrum of engineering applications, ranging from electromagnetic wave manipulation \cite{waveguide, damaskos1982dispersion, itin2010dispersion, kim2023extraordinary} to cancer treatment \cite{jiao2012quantitative, fan2017novel, yao2021biological}. Such materials usually possess highly complex disordered microstructures, which poses great challenges for their design and optimization. Topology optimization, which finds an optimal distribution of material phases on a pre-defined regular grid to achieve targeted material properties, typically via an integrative optimization scheme, is a widely employed numerical design technique for heterogeneous material systems \cite{sigmund1999design, sigmund1997design}. The application of the topology optimization technique in three-dimensional (3D) complex material systems is very limited due to the high computational cost and bad convergence performance associated with the extremely large number of design variables, e.g., $10^6$ to $10^8$ phase voxels. 







An alternative material design approach is to leverage data-driven \cite{xu2022correlation} or analytical \cite{cang2018improving, cheng2022data} structure-property relationship to significantly speed up material property calculations. For example, the strong-contrast expansion formalism expresses the effective material properties \cite{torquato1997effective, torquato1985effective, kim2020effective, torquato2021nonlocal, kim2023effective}  in terms of certain integrals involving the statistical microstructural descriptors, i.e., $n$-point correlation functions $S^{(i)}_n$ of phase $i$ of the material (see Sec. II).  Due to the superior convergence of the strong-contrast formalism, accurate estimates of effective material properties can be obtained by truncating the full series expansions at lower-order correlation functions \cite{torquato1997effective, torquato1985effective}. In the case of both elastic \cite{kim2020effective} and electromagnetic wave propagation \cite{torquato2021nonlocal, kim2023effective} problems, it has been shown that the effective wave properties can be accurately approximated by only employing the {\it spectral density function} ${\tilde \chi}_{_V}({\bf k})$, which is the Fourier transform of the autocovariance function ${\chi}_{_V}({\bf r}) = S_2({\bf r}) - \phi^2$, where $S_2({\bf r})$ and $\phi$ are respectively the two-point correlation function and volume fraction of the phase of interest (see Sec. II for details). ${\tilde \chi}_{_V}({\bf k})$ can be obtained directly from scattering experiments \cite{debye1949scattering, debye1957scattering} and also appears in the rigorous expression of time-dependent diffusion spreadability $\mathcal{S}(t)$ \cite{torquato2021diffusion, skolnick2023simulated} and an approximation scheme for fluid permeability $k$ \cite{torquato1990rigorous, torquato2020predicting}.






Generating realizations of heterogeneous materials with a prescribed set of statistical microstructural descriptors \cite{roberts1997statistical, niezgoda2008delineation, okabe2005pore, jiao2007modeling, jiao2008modeling, hajizadeh2011multiple, tahmasebi2013cross, tahmasebi2012multiple, xu2013stochastic, xu2014descriptor, cang2017microstructure, yang2018microstructural,li2018transfer, farooq2018spectral, cheng2022data, xu2022correlation, skolnick2024quantifying, shih2024fast, casiulis2024gyromorphs} is a crucial inverse problem for material design, which is usually referred to as the microstructure {\it construction} problem \cite{To02a, Ye98a, Ye98b}. One of the most widely used construction methods, among others \cite{roberts1997statistical, gao2021ultra, niezgoda2008delineation, fullwood2008microstructure, cherkasov2021adaptive, okabe2005pore, hajizadeh2011multiple, tahmasebi2013cross, tahmasebi2012multiple, xu2014descriptor, cang2017microstructure, cang2018improving, yang2018microstructural,li2018transfer, cheng2022data}, is the Yeong-Torquato (YT) procedure, in which the construction is formulated as an energy minimization problem \cite{Ye98a, Ye98b}, subsequently solved using simulated annealing \cite{kirkpatrick1983optimization}. The YT procedure has been employed to incorporate a variety of spatial correlation functions \cite{jiao2007modeling, jiao2008modeling, jiao2009superior, chen2019hierarchical, chen2020probing} and applied to a wide spectrum of two-phase material systems \cite{gerke2015improving, karsanina2018hierarchical, feng2018accelerating, jiao2013modeling, chen2015dynamic, jiao2014modeling, guo2014accurate, chen2016stochastic, gerke2019calculation, karsanina2022stochastic, chen2022quantifying}. Given the importance of the spectral density function ${\tilde \chi}_{_V}({\bf k})$ in determining effective material properties, the YT procedure was recently generalized to a Fourier-space based construction framework, which directly incorporates ${\tilde \chi}_{_V}({\bf k})$ as the target function \cite{Ch18a, shi2023computational}. 




In this paper, disordered two-phase heterogeneous materials (media) are of central interest. Among other disordered media, we investigate a class of recently discovered disordered {\it hyperuniform} media \cite{To03, Za09, To16a, To18a}. A hyperuniform medium possesses a vanishing spectral density function in the zero-wavenumber limit, i.e., $\lim_{|{\bf k}|\rightarrow 0}\Tilde{\chi}_{_V}({\bf k}) = 0$. Since $\Tilde{\chi}_{_V}({\bf k})$ is trivially proportional to the scattering intensity \cite{To02a}, this indicates that the scattering of a disordered hyperuniform heterogeneous material is completely suppressed at the infinite-wavelength limit. Equivalently, a hyperuniform heterogeneous material, disordered or not, possesses a local volume fraction variance $\sigma_{_V}^2(R)$ that decreases {\it more rapidly} than $R^d$ for large $R$, i.e., $\lim_{R\rightarrow\infty}\sigma_{_V}^2(R) \cdot R^d = 0$, where $R$ is the radius of spherical observation windows used to compute $\sigma_{_V}^2(R)$. This behavior is to be contrasted with those of typical disordered two-phase media for which the variance decays as $R^{-d}$.


Disordered hyperuniform media are similar to liquids or glasses in that they are statistically isotropic and lack conventional long-range order, and yet they completely suppress large-scale normalized density fluctuations like crystals \cite{To03, Za09, To16a, To18a}. This unique defining feature endows them with many superior physical properties, including wave propagation characteristics \cite{ref31, ref32, ref33, scattering, granchi2022near, park2021hearing, klatt2022wave, tavakoli2022over, cheron2022wave, yu2021engineered, li2018biological}, thermal, electrical and diffusive transport properties \cite{ref34, torquato2021diffusion, maher2022characterization}, mechanical properties \cite{ref35, puig2022anisotropic} as well as optimal multifunctional characteristics \cite{ref36, kim2020multifunctional, torquato2022extraordinary}. We note that besides two-phase media, hyperuniformity has been identified in a wide spectrum of physical \cite{ref4, ref5, ref6, ref7, ref16, ref17, ref18, ref19, ref20,
ref21, ref22, ref23, salvalaglio2020hyperuniform, hexner2017noise, hexner2017enhanced, weijs2017mixing,
lei2019nonequilibrium, lei2019random, ref8, ref9, ref10, ref11,
ref12, ref13, ref14, ref15, ref24, ref25, sanchez2023disordered}, material \cite{ref28, ref29, ref30, Ge19, sakai2022quantum, Ru19, Sa19, Zh20, Ch21, PhysRevB.103.224102, Zh21, nanotube, zhang2023approach, chen2021multihyperuniform} and biological \cite{ref26, ref27, ge2023hidden, liu2024universal, tang2024tunablehyper} systems. The readers are referred to the recent review article by Torquato \cite{To18a} for further details about hyperuniform states of matter.

The construction of {\it isotropic} disordered hyperuniform media from realizable {\it angular-averaged} spectral density functions ${\tilde \chi}_{_V}({k})$ (where $k = |{\bf k}|$) was investigated in detail in Ref. \cite{Ch18a}, including a special class of stealthy hyperuniform materials which possess a spectral density ${\tilde \chi}_{_V}({k}) = 0$ for ${k} < K^*$ \cite{To15}. Ref. \cite{shi2023computational} focused on {\it anisotropic} stealthy hyperuniform composites, which possess ${\tilde \chi}_{_V}({\bf k}) = 0$ for ${\bf k} \in \Omega$, where $\Omega$ is an ``exclusion'' region around the origin in the $\bf k$-space in which scattering is completely suppressed. These results are crucial to engineering hyperuniform media to achieve exotic anisotropic dispersion relations for electromagnetic and acoustic wave propagation \cite{damaskos1982dispersion, itin2010dispersion}. However, both studies focused on constructing heterogeneous material systems in two-dimensions alone.

Here, we present an efficient Fourier-space based computational framework, which further generalizes the techniques developed in Refs. \cite{Ch18a} and \cite{shi2023computational}, to construct {\it three-dimensional} (3D) statistically isotropic two-phase heterogeneous materials corresponding to targeted spectral density functions. This achieved by investigating a variety of analytical ${\tilde \chi}_{_V}({k})$ that satisfy all known necessary conditions characterizing stealthy hyperuniform, standard hyperuniform, nonhyperuniform, and antihyperuniform media. By tuning the correlation lengths in the targeted ${\tilde \chi}_{_V}({k})$, we construct a rich spectrum of distinct microstructures within each class above. We emphasize that the 3D realizations of the majority of ${\tilde \chi}_{_V}({k})$ investigated have never been explicitly constructed before. Importantly, we present the first realization of antihyperuniform two-phase heterogeneous materials in 3D, which are characterized by a power-law autocovariance function $\chi_{_V}(r)$ and contain clusters of dramatically different sizes and morphologies. 



Subsequently, we demonstrate how the effective properties of the constructed materials can be controlled via the targeted ${\tilde \chi}_{_V}({k})$ by estimating the diffusion spreadability ${\cal S}(t)$ and fluid permeability $k$ associated with all of the constructed materials directly from the corresponding ${\tilde \chi}_{_V}({k})$ functions. We show that varying the length-scale parameter within each class of functions can lead to orders of magnitude variation of ${\cal S}(t)$ at intermediate and long time scales, although the long-time asymptotic scaling behavior of ${\cal S}(t)$ only depends on the functional form of ${\tilde \chi}_{_V}({k})$. with the stealthy hyperuniform and antihyperuniform media respectively achieving the most and least efficient transport. In addition, we find that increasing solid volume fraction $\phi_1$ and correlation length $a$ in the constructed media generally leads to a decrease in the dimensionless fluid permeability $k/a^2$, while the antihyperuniform media possess the largest $k/a^2$ among the four classes of materials with the same $\phi_1$ and $a$. These results indicate the feasibility and effectiveness of employing parameterized ${\tilde \chi}_{_V}({k})$ for designing composites with targeted transport properties. 

The rest of the paper is organized as follows: In Sec. II, we provide definition of correlation functions, spectral density function, and hyperuniformity in heterogeneous two-phase material systems, and discuss the effective properties including the diffusion spreadability ${\cal S}(t)$ and fluid permeability $k$. In Sec. III, we discuss the numerical construction procedure in detail, including its mathematical formulation as a constrained optimization problem and its solution via stochastic simulated annealing method. In Sec. IV, we present constructions of realizations of a variety of disordered microstructures with prescribed realizable analytical spectral density functions. In Sec. V, we present the results on ${\cal S}(t)$ and $k$ associated with the constructed materials. In Sec. VI, we provide concluding remarks and outlook of future work.

\section{Definitions}
\label{definition}



\subsection{Correlation Functions and Spectral Density Function}


Here we focus on a two-phase random heterogeneous material (i.e., a random medium), which is a sub-domain of $d$-dimensional Euclidean space, i.e., $\mathcal{V} \subseteq \mathbb{R}^d$ of volume $V \leq +\infty$, composed of two regions $\mathcal{V} = \mathcal{V}_1 \cup \mathcal{V}_2$. $\mathcal{V}_1$ is the phase 1 region with volume $V_1$ and volume fraction $\phi_1 = V_1/V$; and $\mathcal{V}_2$ is the phase 2 region with volume $V_2$ and volume fraction $\phi_2 = V_2/V$. In the infinite-volume limit $V\rightarrow \infty$, $V_i$ ($i=1, 2$) also increases proportionally such that $\phi_i = V_i/V$ tends to a well-defined constant. The statistical properties of each phase $i$ of the system are specified by the countably infinite set of {\it $n$-point correlation functions} $S_n^{(i)}$, which are defined by
\cite{To02a}:
\begin{equation}\label{Sndef}
S_n^{(i)}(\mathbf{x}_1, \ldots, \mathbf{x}_n) = \left\langle\prod_{i=1}^n I^{(i)}(\mathbf{x}_i)\right\rangle,
\end{equation}
where $I^{(i)}({\bf x})$ is the indicator function for phase $i$, i.e.,
\begin{equation}
I^{(i)}({\bf x}) =
\begin{cases}
1, \quad &\text{if } {\bf x} \in  \mathcal{V}_1\\
0, \quad &\text{otherwise.}
\end{cases}
\end{equation}
The function $S_n^{(i)}(\mathbf{x}_1, \ldots, \mathbf{x}_n)$ can also be interpreted as the probability of finding a set of randomly selected $n$ points at positions $\mathbf{x}_1, \ldots, \mathbf{x}_n$ all falling into the same phase $i$. Henceforth, we consider statistically homogeneous systems, i.e., there is no preferred origin in the system and thus $S_n^{(i)}$ only depends on the relative displacements between the points \cite{torquato2002random}.

The one-point function is simply the volume fraction of phase $i$, $\phi_i$, which is
a constant, i.e.,
\begin{equation}
S_1^{(i)}(\mathbf{x_1}) =  \phi_i,
\end{equation}
and the two-point correlation function $S_2^{(i)}({\bf r})$ depends only on the relative displacement ${\bf r}={\bf x}_2-{\bf x}_1$. For heterogeneous materials without long-range order, $S_2({\bf r})$ possesses the following asymptotic behavior:
\begin{equation}
\lim_{|{\bf r}|\rightarrow \infty} S^{(i)}_2(\mathbf{r}) = \phi_i^2.
\end{equation}


Upon subtracting the long-range
behavior from $S^{(i)}_2$, one obtains the {\it autocovariance function}, i.e.,
\begin{equation}
\chi_{_V}({\bf r})= S^{(i)}_2(\mathbf{r})-\phi_i^2
\end{equation}
which is generally an $L^2$-function. The associated spectral density
function $\Tilde{\chi}_{_V}({\bf k})$ is given by
\begin{equation}
\Tilde{\chi}_{_V}({\bf k}) = \int_{\mathbb{R}^d} \chi_{_V}({\bf r})e^{-i{\bf
k}\cdot{\bf r}}d{\bf r},
\end{equation}
which is the Fourier transform of $\chi_{_V}({\bf r})$ and is obtainable experimentally from scattering intensity measurements \cite{debye1949scattering, debye1957scattering}. For statistically isotropic two-phase random media, i.e., the focus of this work, the autocovariance function only depends on the distance $r = |{\bf r}|$ between the two points, i.e., $\chi_{_V}({r}) = \chi_{_V}(|{\bf r}|)$. Accordingly, the spectral density function only depends on the wavenumber $k = |{\bf k}|$, i.e., $\Tilde{\chi}_{_V}({k}) = \Tilde{\chi}_{_V}({\bf k})$.

We note that the autocovariance function obeys the bounds \cite{torquato2006necessary}:
\begin{equation}
-\text{min}\{(1-\phi)^2, \phi^2\} \leq \chi_{_V}({\bf r})\leq (1-\phi)\phi,
\end{equation}
where $\phi$ is the volume fraction of the reference phase.
We remark that it is an open problem to identify additional necessary and sufficient conditions \cite{torquato2006necessary} that the autocovariance function must satisfy in order to correspond to a binary stochastic process. 


\subsection{Hyperuniform, Nonhyperuniform and Antihyperuniform Random Media}


In the context of a hyperuniform two-phase heterogeneous material, the quantity of interest is the local volume fraction variance $\sigma_{_V}^2(R)$, which was first introduced in Ref. \cite{lu1990local}:
\begin{equation}
\sigma_{_V}^2(R) = \frac{1}{v_1(R)}\int_{\mathbb{R}^d} I({\bf r})\alpha_2(r; R)d{\bf r},
\end{equation}
where $\alpha_2(r; R)$ is the scaled intersection volume, i.e., the intersection volume of two spherical windows of radius $R$ whose centers are separated by a distance $r$, divided by the volume $v_1(R)$ of the window, i.e.,
\begin{equation}
v_1(R)=\frac{\pi^{d/2} R^d}{\Gamma(1+d/2)}.
\label{v1}
\end{equation}




A disordered hyperuniform medium is one whose
$\sigma_{_V}^2(R)$ decreases more rapidly than $R^d$ for large $R$ \cite{Za09}, i.e.,
\begin{equation}
\lim_{R\rightarrow\infty}\sigma_{_V}^2(R) \cdot R^d = 0.
\end{equation}
This behavior is to be contrasted with those of typical disordered two-phase media for which the variance decays as $R^{-d}$, i.e., as the inverse of the window volume $v_1(R)$.


The hyperuniform condition is equivalently given by
\begin{equation}
\label{eq_hyper} \lim_{|{\bf k}|\rightarrow 0}\Tilde{\chi}_{_V}({\bf k})
= 0,
\end{equation}
which implies that  the  direct-space autocovariance function $\chi_{_V}({\bf r})$  exhibits both positive  and  negative  correlations  such  that  its volume  integral  over all space is exactly zero \cite{To16b}, i.e.,
\begin{equation}
\int_{\mathbb{R}^d} \chi_{_V}({\bf r})d{\bf r} = 0.
\label{eq_sum_rule}
\end{equation}
Eq. (\ref{eq_sum_rule}) is a direct-space sum rule for hyperuniformity. 


For hyperuniform two-phase media whose spectral density goes to zero as a power-law scaling as $|\bf k|$ tends to zero \cite{To16a}, i.e.,
\begin{equation}
    {\tilde \chi}_{_V}({\bf k})\sim |{\bf k}|^\alpha.
\end{equation}
Stealthy hyperuniform media are those in which the spectral density function is exactly zero for a range of wavevectors around the origin \cite{To18a, torquato2021diffusion}, i.e.,
\begin{equation}
    {\tilde \chi}_{_V}({\bf k}) = 0, \quad {\bf k} \in \Omega,
\end{equation}
where $\Omega$ is a finite region around the origin of the Fourier space. It is also convenient to introduce a ratio $\chi$ between the number of constraints $N_\Omega$ and the number of degrees of freedom $N$ for the stealthy hyperuniform systems:
\begin{equation}
    \chi = N_\Omega/(N-3)
    \label{eq_chi_ratio}
\end{equation}
where 3 degrees of freedom associated with the trivial overall translation of the entire system are subtracted in Eq. (\ref{eq_chi_ratio}).

There are three different scaling regimes (classes) that describe the associated large-$R$ behaviors of the local volume fraction variance:
\begin{equation}
\sigma^2_{_V}(R) \sim
\begin{cases}
R^{-(d+1)}, \quad\quad\quad \alpha >1 \qquad &\text{(Class I)}\\
R^{-(d+1)} \ln R, \quad \alpha = 1 \qquad &\text{(Class II)}\\
R^{-(d+\alpha)}, \quad 0 < \alpha < 1\qquad  &\text{(Class III).}
\end{cases}
\label{eq:classes}
\end{equation}
Classes I and III are the strongest and weakest forms of hyperuniformity, respectively. Class I media include all crystal structures \cite{To03}, many quasicrystal structures \cite{Og17} and exotic disordered media \cite{Za09, Ch18a}. Examples of Class II systems include some quasicrystal structures \cite{Og17}, perfect glasses \cite{zhang2017classical}, and maximally random jammed packings \cite{ref4, ref5, ref6, Za11c, Za11d}. Examples of Class III systems include classical disordered ground states \cite{Za11b}, random organization models \cite{ref20}, perfect glasses \cite{zhang2017classical}, and perturbed lattices \cite{Ki18a}; see Ref. \cite{To18a} for a more comprehensive list of systems that fall into the three hyperuniformity classes. 


By contrast, for any nonhyperuniform two-phase system, the local volume fraction variance has the following large-$R$ scaling behaviors \cite{torquato2021diffusion}:
\begin{equation}
\sigma^2_{_V}(R) \sim
\begin{cases}
R^{-d}, ~\alpha =0 \quad \text{(standard nonhyperuniform)}\\
R^{-(d+\alpha)}, ~-d < \alpha < 0\quad  \text{(antihyperuniform).}
\end{cases}
\label{eq:classesnon}
\end{equation}
A {\it standard nonhyperuniform} two-phase medium \cite{torquato2021diffusion} is one whose spectral density function is bounded and approaches a finite constant as $|{\bf k}|$ goes to zero, i.e., 
\begin{equation}
   \lim_{|{\bf k}|\rightarrow 0} {\tilde \chi}_{_V}({\bf k})\sim const.
\end{equation}
Examples of standard nonhyperuniform systems include overlapping systems with Poisson distribution of centers, equilibrium hard-sphere fluids, and hard-sphere packings generated via random sequential addition process \cite{To18a, torquato2021local}. An {\it antihyperuniform} two-phase medium \cite{torquato2021diffusion} possesses an unbounded spectral density function in the zero-$|{\bf k}|$ limit, i.e., 
\begin{equation}
   \lim_{|{\bf k}|\rightarrow 0} {\tilde \chi}_{_V}({\bf k})\rightarrow +\infty
\end{equation}
Systems at critical point containing macroscopic scale fluctuations possess a diverging spectral density at the origin and thus are antihyperuniform. Other examples include systems generated via hyperplane intersection process (HIP) and Poisson cluster process \cite{torquato2021local}.

\subsection{Time-dependent Diffusion Spreadability}


The diffusion spreadability $\mathcal{S}(t)$ \cite{torquato2021diffusion} is a dynamical probe that directly links the time-dependent diffusive transport behavior with the microstructure of heterogeneous materials across length and time scales. $\mathcal{S}(t)$ specifically considers the time-dependent mass transfer in a two-phase material where all solute is initially concentrated in phase 2, and the solute has the same diffusion coefficient $D$ in each phase. The spreadability  is therefore
defined as the total solute present in phase 1 at time. A system with larger $\mathcal{S}(t)$ for a given time $t$ can spread diffusion information more efficiently. For a two-phase heterogeneous material in $\mathbb{R}^d$, the spreadability is given by 
\begin{equation}
\begin{array}{c}
\mathcal{S}(t) - \mathcal{S}(\infty) = \displaystyle \frac{1}{(4\pi Dt)^{d/2}\phi_2} \int_{\mathbb{R}^d} \chi_{_V}({\bf r})e^{-|{\bf r}|^2/(4Dt)}d{\bf r} \\
= \displaystyle \frac{1}{(2\pi)^d \phi_2} \int_{\mathbb{R}^d} \tilde{\chi}_{_V}({\bf k})e^{-|{\bf k}|^2 Dt}d{\bf k}
\end{array}
\label{eq_St}
\end{equation}
where $\phi_2$ is the volume fraction of phase 2, $\mathcal{S}(\infty) = \phi_1$ is the infinite-time limit of $\mathcal{S}(t)$, and $\mathcal{S}(t) - \mathcal{S}(\infty)$ is referred to as the excess spreadability. 

We note that Eq.~(\ref{eq_St}) is a rare example of transport in two-phase random materials that is exactly determined by only the first two correlation functions, i.e., $\phi_i$ and $\chi_{_V}({\bf r})$ (or equivalently $\Tilde{\chi}_{_V}({\bf k})$). Based on Eq.~(\ref{eq_St}), Torquato \cite{torquato2021diffusion} demonstrated that the small-, intermediate-, and long-time behaviors of $\mathcal{S}(t)$ are directly determined respectively by the small-, intermediate-, and large-scale structural features of the two-phase material. In particular, for statistically homogeneous microstructures whose spectral densities
possess the following small-$|k|$ scaling behavior:
\begin{equation}
{\tilde \chi}_{_V}({\bf k}) \sim B |{\bf k}|^\alpha,
\end{equation}
Torquato showed that the long-time behavior of excess
spreadability for two-phase media in $\mathbb{R}^d$ is given by the inverse power law:
\begin{equation}
\mathcal{S}(t) - \mathcal{S}(\infty) \sim \frac{B \Gamma[(d+\alpha)/2]\phi_2}{2^d \pi^{d/2} \Gamma(d/2) (Dt/a^2)^{(d+\alpha)/2}}
\label{eq_St_power}
\end{equation}
where $a$ is a characteristic heterogeneity length-scale (see Sec. IV for details) and $B$ is a microstructure-dependent coefficient. The power law (\ref{eq_St_power}) holds except for the case where $\alpha \rightarrow \infty$, which, roughly speaking, may be regarded to correspond to stealthy hyperuniform materials, whose excess spreadability decays exponentially fast, i.e.,
\begin{equation}
\mathcal{S}(t) - \mathcal{S}(\infty) \sim C \exp(-\theta t)/t
\end{equation}
for disordered stealthy hyperuniform two-phase media, and
\begin{equation}
\mathcal{S}(t) - \mathcal{S}(\infty) \sim C \exp(-\theta t)
\end{equation}
for ordered stealthy hyperuniform two-phase media, where $C$ and $\theta$ are microstructure dependent parameters. Therefore, the spreadability is a powerful dynamic probe to categorize all statistically homogeneous two-phase microstructures that span from hyperuniform to nonhyperuniform materials across length scales. In Sec. IV, we will employ Eq.~(\ref{eq_St}) to compute and investigate $\mathcal{S}(t)$ for the wide spectrum of the constructed 3D microstructures. 

\subsection{Fluid Permeability}


A key transport property of two-phase random heterogeneous materials (such as porous media) is the fluid permeability $k$, which is described by the so-called Darcy law \cite{torquato2002random},
\begin{equation}
{\bf U}({\bf x}) = -\frac{k}{\mu} \nabla p_0({\bf x}),
\end{equation}
where ${\bf U}({\bf x})$ is the average fluid velocity, $\nabla p_0({\bf x})$ is the applied pressure gradient, and $\mu$ is the dynamic viscosity. The permeability $k$, which has dimensions of (length)$^2$, depends upon the details of the material microstructure (e.g., pore morphology) in a complex fashion \cite{torquato2002random}. Physically $k$ may be interpreted as an effective cross-sectional area of pore ``channels'' in porous media.

The two-point ``void'' upper bound on the fluid permeability of a general statistically isotropic porous medium in $\mathbb{R}^3$ with porosity $\phi_1 = 1 - \phi_2$ is given by \cite{rubinstein1989flow, torquato2020predicting}
\begin{equation}
k \le \frac{2}{3\phi_2^2} \ell_p^2,
\label{eq_k_bound}
\end{equation}
where $\ell_p$ is a length scale involving the autocovariance function:
\begin{equation}
\ell_p^2 = \int_0^\infty \chi_{_V}(r)rdr = \frac{1}{2\pi^2} \int_0^\infty \tilde{\chi}_{_V}(k)dk
\label{eq_lp}
\end{equation}
Importantly, it has been shown \cite{torquato2020predicting} that bound (\ref{eq_k_bound}) can be employed to construct accurate estimate of $k$ for a porous medium, given that the fluid permeability $k_0$ of a reference microstructure with the same porosity is known (either analytically or from accurate numerical simulations), i.e., 
\begin{equation}
k \approx \frac{\ell_p^2}{\ell_{p0}^2} k_0
\label{eq_k_est}
\end{equation}
where $\ell_{p0}^2$ is the length scale parameter (\ref{eq_lp}) of the reference microstructure. In the subsequent discussions, we will employ Eq. (\ref{eq_k_est}) to estimate the fluid permeability of the constructed 3D microstructures.  


\section{Methods}

\subsection{Formulation of the Construction Problem}


The {\it construction} process aims to find a digital volume (e.g., represented by a binary-valued array with entries equal to 0 or 1) associated with a prescribed set of statistical descriptors. Here, we focus on the construction of a three-dimensional disordered two-phase medium with a prescribed, realizable spectral density $\tilde \chi_{_V}({\bf k})$. Specifically, we consider digitized media in a cubic domain of length $L$ in $\mathbb{R}^3$ with periodic boundary conditions. In this case, the indicator function $I({\bf r})$ takes a discrete form, i.e., $I({\bf r})$ is only defined on a discrete set of ${\bf r} = n_1 {\bf e}_1 + n_2 {\bf e}_2 + n_3 {\bf e}_3$, where ${\bf e}_i$ are unit vectors along the orthogonal directions and $n_1, n_2, n_3 = 0, 1, \ldots, L$ with $L$ being the linear system size or ``resolution'' (i.e., number of voxels along each dimension). The Fourier-space wavevectors for the system also take discrete values, i.e., ${\bf k} = (2\pi/L)(n_1 {\bf e}_1 + n_2 {\bf e}_2 + n_3 {\bf e}_3)$. The corresponding spectral density for the digital medium is given by
\begin{equation}
    \tilde \chi_{_V}({\bf k}) = \frac{1}{L^2}{\tilde m^2}({\bf k})|\tilde J (\bf k)|^2
    \label{eq_chi_k}
\end{equation}
where $\tilde J (\bf k)$ is the {\it generalized collective coordinate} \cite{To18a, Ch18a} defined as
\begin{equation}
 \tilde J (\bf k) = \sum_{{\bf r}}\exp(i{\bf k}\cdot{\bf r}) J({\bf r})
   \label{eq_I_k}
\end{equation}
where the sum is over all pixel centers ${\bf r}$, and
\begin{equation}
J({\bf r}) = I({\bf r}) - \phi.
   \label{eq_J}
\end{equation}
The quantity ${\tilde m}({\bf k})$ is the Fourier transform
of the shape function (or indicator function) $m({\bf r})$ of a square pixel which is given by
\begin{equation}
{\tilde m}({\bf k}) = \Pi_{l=1}^3 \sin(\pi n_l/L)
\end{equation}
where $k_l = \pi n_l/L$ ($l = 1, 2, 3$) are the component of the wavevector ${\bf k}$. 



The construction of a random medium with a given target $\tilde \chi_{_V}^0({\bf k})$ for wavevectors ${\bf k}\in \Omega$ directly imposes a set of constraints on the discrete indicator function $I({\bf r})$ through Eqs. (\ref{eq_chi_k}) to (\ref{eq_J}), i.e.,
\begin{equation}
    {\tilde m^2}({\bf k})|\sum_{\bf r}\exp(i{\bf k}\cdot{\bf r})[I({\bf r})-\phi)]|^2 - \tilde \chi_{_V}^0({\bf k}) = 0
    \label{eq_chi_Ir}
\end{equation}
for ${\bf k}\in \Omega$. We note Eq. (\ref{eq_chi_Ir}) represents a set of $N_\Omega$ number of nonlinear equations of $I({\bf r})$ (and equivalently of $J({\bf r})$), where $N_\Omega$ is the number of {\it independent} $\bf k$ points in $\Omega$. In the case of statistically isotropic systems, $\Omega$ corresponds to a spherical region in the wavevector space. We note that due to the symmetry of the spectral density function $\tilde \chi_{_V}({\bf k})$ \cite{Ch18a}, only a half of ${\bf k}$ points in $\Omega$ are independent and thus, $N_\Omega \sim \frac{1}{2}Vol(\Omega)$. For a 3D digital realization of linear size $L$, the total number of voxels is $N = L^3$. The number of unknowns in Eq. (\ref{eq_chi_Ir}), i.e., the value of each voxel, is also $N$. We are interested in the cases $N_\Omega < N$, i.e., the number of constraints (equations) is smaller than that of unknowns. Therefore, Eq. (\ref{eq_chi_Ir}) does not have unique solutions and we will employ stochastic optimization method to iteratively solve Eq. (\ref{eq_chi_Ir}).






\subsection{Simulated Annealing Procedure}


We employ the simulated annealing procedure \cite{kirkpatrick1983optimization} to solve Eq. (\ref{eq_chi_k}), which has been widely used in material construction problems \cite{Ye98a, Ye98b, jiao2007modeling, jiao2008modeling, jiao2009superior, jiao2013modeling, jiao2014modeling, Ch18a, shi2023computational}. In particular, the construction problem is formulated as
an ``energy'' minimization problem, with the energy functional $E$ defined as follows
\begin{equation}
\label{eq208} E =
\sum\limits_{{\bf k}\in \Omega}\left[{\tilde \chi_{_V}({\bf k})-\tilde \chi_{_V}^{0}({\bf k})}\right]^2,
\end{equation}
where $\tilde \chi_{_V}^{0}({\bf k})$ is the target spectral density function and $\tilde \chi_{_V}({\bf k})$ is the corresponding function associated with a trial microstructure, and $\Omega$ is the set of constrained {\it independent} ${\bf k}$ points. For statistically isotropic two-phase random media, i.e., the focus of this work, the spectral density function only depend on the wavenumber $k = |{\bf k}|$. Thus, the corresponding energy functional is given by 
\begin{equation}
E =
\sum\limits_{k < K^*}\left[{\tilde \chi_{_V}({k})-\tilde \chi_{_V}^{0}({k})}\right]^2,
\end{equation}
where $K^*$ is the upper bound on the constrained wavenumbers, and $\tilde \chi_{_V}({k})$ can be obtained from $\tilde \chi_{_V}({\bf k})$ by taking angular averages.




The aforementioned energy minimization problem is then solved using simulated annealing \cite{kirkpatrick1983optimization}. Specifically, starting from
an initial trial configuration (i.e., old realization) with an energy $E_{old}$, which
contains a fixed number of pixels for each phase (specified by the corresponding volume fraction $\phi_i$), two randomly selected pixels
associated with different phases are switched to generate a new
trial microstructure. The new spectral density function is sampled
from the new trial configuration and the associated energy $E_{new}$ is
evaluated, which determines whether the new trial configuration
should be accepted or rejected via the probability \cite{kirkpatrick1983optimization}:
\begin{equation}
\label{eq_pacc} p_{acc} = min \{\exp(-\Delta E/T), 1\},
\end{equation}
where $\Delta E = E_{new} - E_{old}$ is the energy difference between the new and old
trial configurations and $T$ is a virtual temperature that is
chosen to be initially high and slowly decreases according to a
cooling schedule. An appropriate cooling
schedule reduces the possibility that the system gets stuck in a
shallow local energy minimum. In practice, a power law schedule
$T(n) = \gamma^n T_0$ is usually employed, where $T_0$ is the
initial temperature, $n$ is the cooling stage and $\gamma \in (0,
1)$ is the cooling factor ($\gamma = 0.99$ is used here). The
simulation is terminated when $E$ is smaller than a prescribed
tolerance (e.g., $10^{-12}$ in this work).

As a stochastic optimization technique, simulation annealing generally needs to search a large number of trial configurations ($\sim 10^7$) in order to generate a successful construction. Therefore, highly efficient
methods \cite{Ch18a, shi2023computational} have been employed to
rapidly obtain the spectral density function $\tilde \chi_{_V}({k})$ of a new configuration by updating the corresponding function associated
with the old configuration, instead of completely re-computing
the function from scratch. Specifically, the generalized collective coordinate $\tilde J({\bf k})$ is tracked: At the
beginning of the construction, $\tilde J({\bf k})$ of the initial configuration is computed from scratch and the values for all ${\bf k}$ are stored. After each voxel switch, since only a single voxel of the phase of interest is moved from ${\bf r}_{old}$ to ${\bf r}_{new}$, the updated $\tilde J({\bf k})$ values can be obtained by only explicitly computing the contributions from this changed voxel, i.e.,
\begin{equation}
\tilde J({\bf k}) \leftarrow \tilde J({\bf k}) + \delta \tilde J_{new}({\bf k}) - \delta \tilde J_{old}({\bf k}),
\end{equation}
where
\begin{equation}
\delta \tilde J_{new}({\bf k}) = \exp(i{\bf k}\cdot {\bf r}_{new}), \quad \delta \tilde J_{old}({\bf k}) = \exp(i{\bf k}\cdot {\bf r}_{old})
\end{equation}
Once the updated $\tilde J({\bf k})$ is obtained, the updated $\tilde \chi_{_V}({\bf k})$ can be immediately computed using Eq. (\ref{eq_chi_k}).


In our subsequent constructions, we mainly consider realizations in a cubic domain with $L = 128$ pixels and periodic boundary conditions. We have also investigated smaller and larger system sizes, including $L = 64$ and $L = 256$ pixels to verify that a resolution of $L = 128$
pixels does not affect the construction results.


\section{Construction of Nonhyperuniform, Hyperuniform and Antihyperuniform Two-Phase Media}

In this section, we present construction results of 3D
statistically isotropic two-phase random media, including
nonhyperuniform, hyperuniform and antihyperuniform ones. The different systems are modeled by their respective parameterized spectral density function $\tilde \chi_{_V}({k}; a)$, where $a$ is a length scale parameter controlling the microstructure within each class of functions (see detailed discussion below). For each selected length scale parameter, we consider three different volume fractions for the phase of interest, i.e., $\phi = 0.25$, 0.5 and 0.75. These selections of $(a, \phi)$ allow us to explore and generate a wide spectrum of distinct microstructures for subsequent physical property optimization. The analytical $\tilde \chi_{_V}({k}; a)$ we consider here satisfy all known necessarily conditions and have been employed to model transport properties \cite{torquato2021diffusion}. However, the 3D realizations of the majority of these functions have never been explicitly obtained before.     


\subsection{Standard Nonhyperuniform Two-Phase Media}

\begin{figure}[ht]
\begin{center}
$\begin{array}{c}\\
\includegraphics[width=0.47\textwidth]{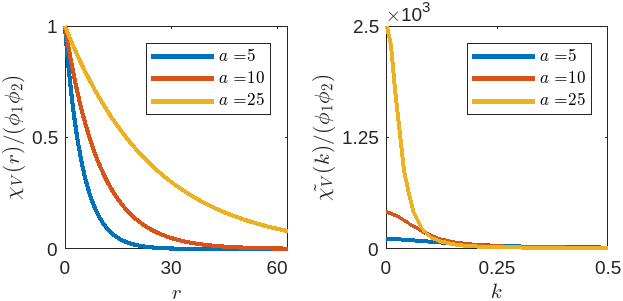}
\end{array}$
\end{center}
\caption{Rescaled autocovariance function $\chi_{_V}({r}; a)/(\phi_1\phi_2)$ (left) and rescaled spectral density function $\tilde \chi_{_V}({k}; a)/(\phi_1\phi_2)$ (right) associated with the nonhyperuniform Debye random media with varying length scale parameter $a$. The unit of $r$ is the edge length of a single voxel and the unit of $k$ is $2\pi/L$.} \label{fig_chi_Debye}
\end{figure}

\begin{figure}[ht]
\begin{center}
$\begin{array}{c}\\
\includegraphics[width=0.49\textwidth]{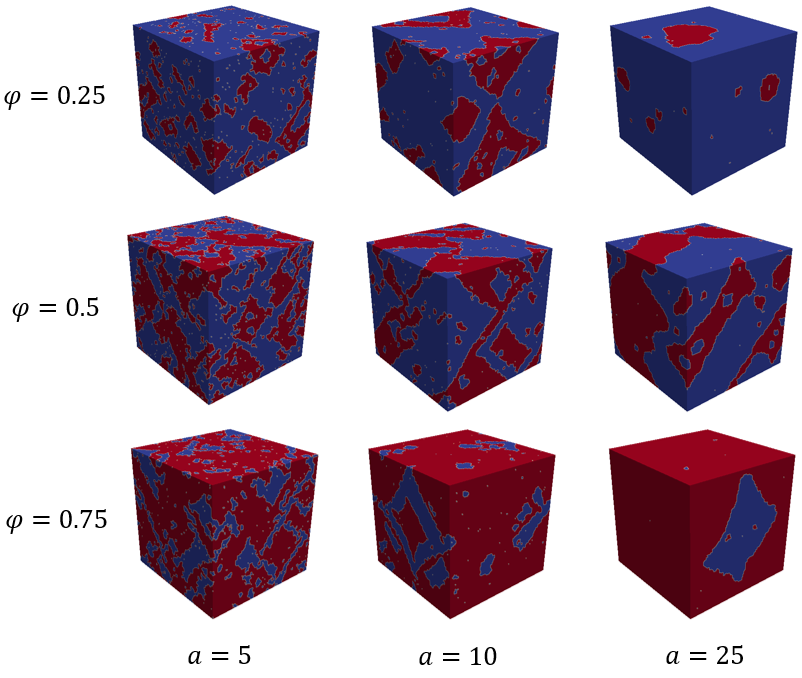}
\end{array}$
\end{center}
\caption{Constructions of the Debye random media with $\phi_1 = 0.25$, 0.5 and 0.75 (top to bottom) and $a = 5$, 10 and 25 voxels, where phase 1 is shown in red color. The linear size of the system is $L = 128$ voxels.} \label{fig_config_Debye}
\end{figure}

We first consider an example of standard nonhyperuniform two-phase medium, i.e., the Debye random medium \cite{debye1957scattering}, which possesses an exponentially decaying autocovariance function, i.e., 
\begin{equation}
\chi_{_V}({r}; a) = \phi_1 \phi_2 \exp(-r/a),
\end{equation}
where $\phi_i$ is the volume fraction of phase $i$, and $a$ is the length scale parameter, i.e., the correlation length. The associated spectral density function $\tilde \chi_{_V}({k})$ is given by \cite{torquato2021diffusion}
\begin{equation}
\tilde \chi_{_V}({k}; a) = \phi_1 \phi_2 \frac{4a^2}{[1+(ka)^2]^{3/2}}.
\label{eq_chi_Debye}
\end{equation}
It follows immediately from Eq. (\ref{eq_chi_Debye}) that 
\begin{equation}
\lim_{ k\rightarrow 0} \tilde \chi_{_V}({k}; a) = 4\phi_1\phi_2 a^2
\end{equation}
which is a non-zero positive constant for any non-zero $a$.
Although this system has been extensively investigated \cite{ma2020generation, skolnick2021understanding}, the previous constructions were based on the autocovariance function $\chi_{_V}({r})$, instead of the spectral density function $\tilde \chi_{_V}({k})$. In particular, it has been shown that realizations of the Debye random medium contain phases of ``fully random shape, size, and distribution.'' Therefore, the Debye random medium provides an excellent bench-mark system to validate the Fourier-space-based 3D construction procedure. 

Figure \ref{fig_chi_Debye} shows the rescaled autocovariance function $\chi_{_V}({r}; a)/(\phi_1\phi_2)$ (left) and rescaled spectral density function $\tilde \chi_{_V}({k}; a)/(\phi_1\phi_2)$ (right) associated with the Debye random media with varying length scale parameter $a = 5$, 10 and 25 voxels. It is apparent that increasing $a$ leads to a longer ranged correlation function, which is manifested as increasing cluster size in the constructed realizations shown in Fig. \ref{fig_config_Debye}. Specifically, a clear growth of the random clusters associated with increasing $a$ can be observed, across all volume fractions $\phi_1$ considered. On the other hand, increasing $\phi_1$ leads to an expected percolation of phase 1 around $\phi_1$ across different $a$ values. Interesting, we note the constructed microstructures with the same correlation length $a$ exhibit the so-called phase inversion symmetry \cite{torquato2002random, shi2023computational}, i.e., the morphology of phase 1 at volume fraction $\phi_1$ is statistically identical to that of phase 2 in the system where the volume fraction is $\phi_2 = 1- \phi_1$. This is because these phases possess identical autocovariance function by construction. These results are consistent with previous $\chi_{_V}({r})$-based construction studies \cite{ma2020generation, skolnick2021understanding} and clearly demonstrate the validity of our Fourier-space-based 3D construction procedure. 



\subsection{Standard Hyperuniform Two-Phase Media}

\begin{figure}[ht]
\begin{center}
$\begin{array}{c}\\
\includegraphics[width=0.47\textwidth]{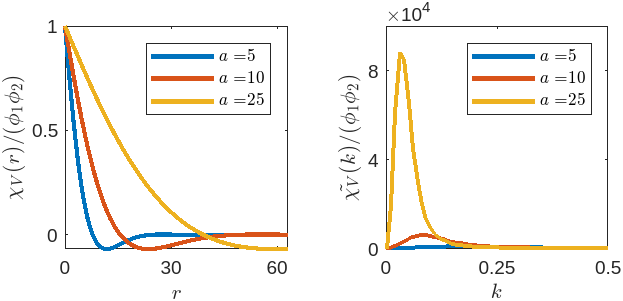}
\end{array}$
\end{center}
\caption{Rescaled autocovariance function $\chi_{_V}({r}; a)/(\phi_1\phi_2)$ (left, cf. Eq. (\ref{eq_auto_std})) and rescaled spectral density function $\tilde \chi_{_V}({k}; a)/(\phi_1\phi_2)$ (right, cf. Eq. (\ref{eq_chi_std})) associated with the standard hyperuniform media with varying length scale parameter $a$. The unit of $r$ is the edge length of a single voxel and the unit of $k$ is $2\pi/L$.} \label{fig_chi_std}
\end{figure}

\begin{figure}[ht]
\begin{center}
$\begin{array}{c}\\
\includegraphics[width=0.49\textwidth]{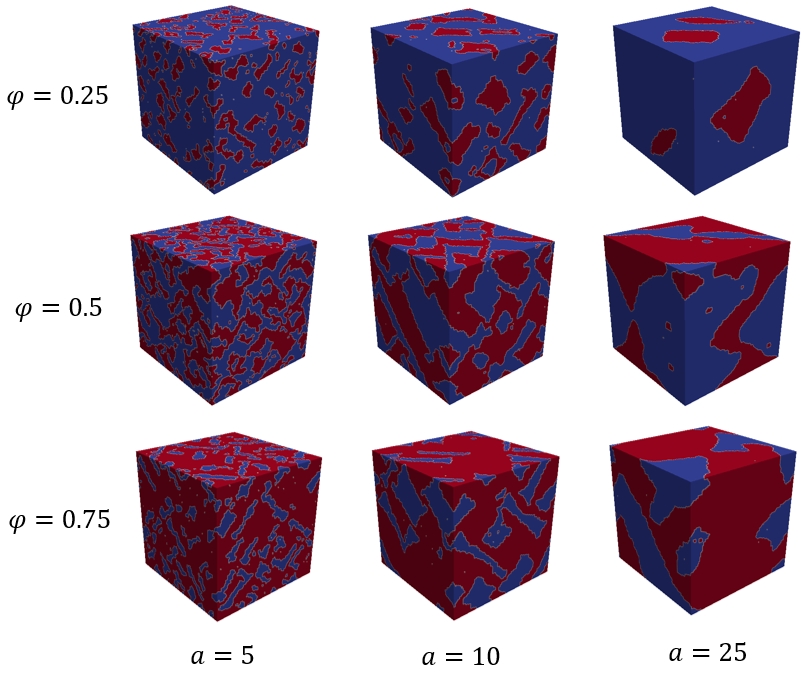}
\end{array}$
\end{center}
\caption{Constructions of the standard hyperuniform media with $\phi_1 = 0.25$, 0.5 and 0.75 (top to bottom) and $a = 5$, 10 and 25 voxels, where phase 1 is shown in red color. The linear size of the system is $L = 128$ voxels.} \label{fig_config_std}
\end{figure}

We now consider a representative standard hyperuniform two-phase medium, which possesses a damped oscillating autocovariance function, i.e., 
\begin{equation}
\chi_{_V}({r}; a) = \phi_1 \phi_2 \exp(-r/a) \cos(qr),
\label{eq_auto_std}
\end{equation}
where $\phi_i$ is the volume fraction of phase $i$, $a$ is the correlation length and $(qa)^2 = 1/3$ \cite{torquato2021diffusion}, which satisfies the hyperuniformity condition (\ref{eq_sum_rule}). The associated spectral density function $\tilde \chi_{_V}({k})$ is given by 
\begin{equation}
\tilde \chi_{_V}({k}; a) = \frac{\phi_1 \phi_2  216 \pi [3(ka)^2 + 8](ka)^2 a^3}{81(ka)^8 + 216(ka)^6 + 432(ka)^4 + 384(ka)^2 + 256}.
\label{eq_chi_std}
\end{equation}
From Eq. (\ref{eq_chi_std}), it can be seen that the spectral density of the system possesses a small-$k$ scaling
\begin{equation}
\tilde \chi_{_V}({k}; a) \sim k^2,
\end{equation}
which belongs to class-I hyperuniformity. 


Figure \ref{fig_chi_std} shows the rescaled autocovariance function $\chi_{_V}({r}; a)/(\phi_1\phi_2)$ (left) and rescaled spectral density function $\tilde \chi_{_V}({k}; a)/(\phi_1\phi_2)$ (right) associated with this standard hyperuniform media with varying length scale parameter $a = 5$, 10 and 25 voxels. Since the correlation length $a$ is coupled with the spatial periodicity $1/q$ through $(qa)^2 = 1/3$ to ensure hyperuniformity, it can be seen that increasing the correlation length $a$ also leads to an increase of the spatial periodicity or equivalently a decrease of the oscillation frequency. 

Figure \ref{fig_config_std} shows the spectrum of microstructures associated with varying $\phi_1$ and $a$. At $\phi_1 = 0.25$, the phase forms well-defined local clusters, possessing a mean size determined by $a$ and a narrow size distribution. The spatial distribution of such clusters also exhibits a very uniform pattern. These features together give rise to the overall hyperuniformity in the system. Increasing $\phi_1$ to 0.5 leads to the percolation of the local clusters to form connected ligaments, whose mean width is again mainly determined by and increases with $a$. A phase inversion symmetry is observed, i.e., at $\phi_1 = 0.75$, the microstructure with a specific $a$ is statistically equivalent to that with the same $a$ and $\phi_1 = 0.25$, but with the red and blue phases exchanged. 




\subsection{Stealthy Hyperuniform Two-Phase Media}

\begin{figure}[ht]
\begin{center}
$\begin{array}{c}\\
\includegraphics[width=0.47\textwidth]{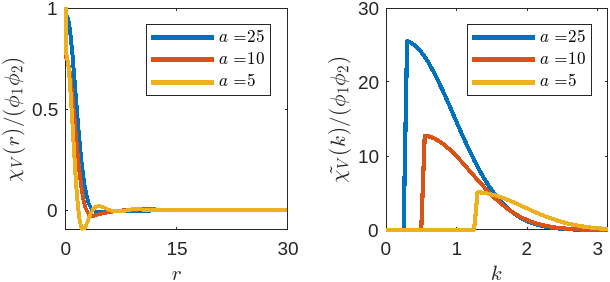}
\end{array}$
\end{center}
\caption{Rescaled autocovariance function $\chi_{_V}({r}; a)/(\phi_1\phi_2)$ (left) and rescaled spectral density function $\tilde \chi_{_V}({k}; a)/(\phi_1\phi_2)$ (right) associated with the stealthy hyperuniform media with varying length scale parameter $a$. The unit of $r$ is the edge length of a single voxel and the unit of $k$ is $2\pi/L$.} \label{fig_chi_stealthy}
\end{figure}

\begin{figure}[ht]
\begin{center}
$\begin{array}{c}\\
\includegraphics[width=0.49\textwidth]{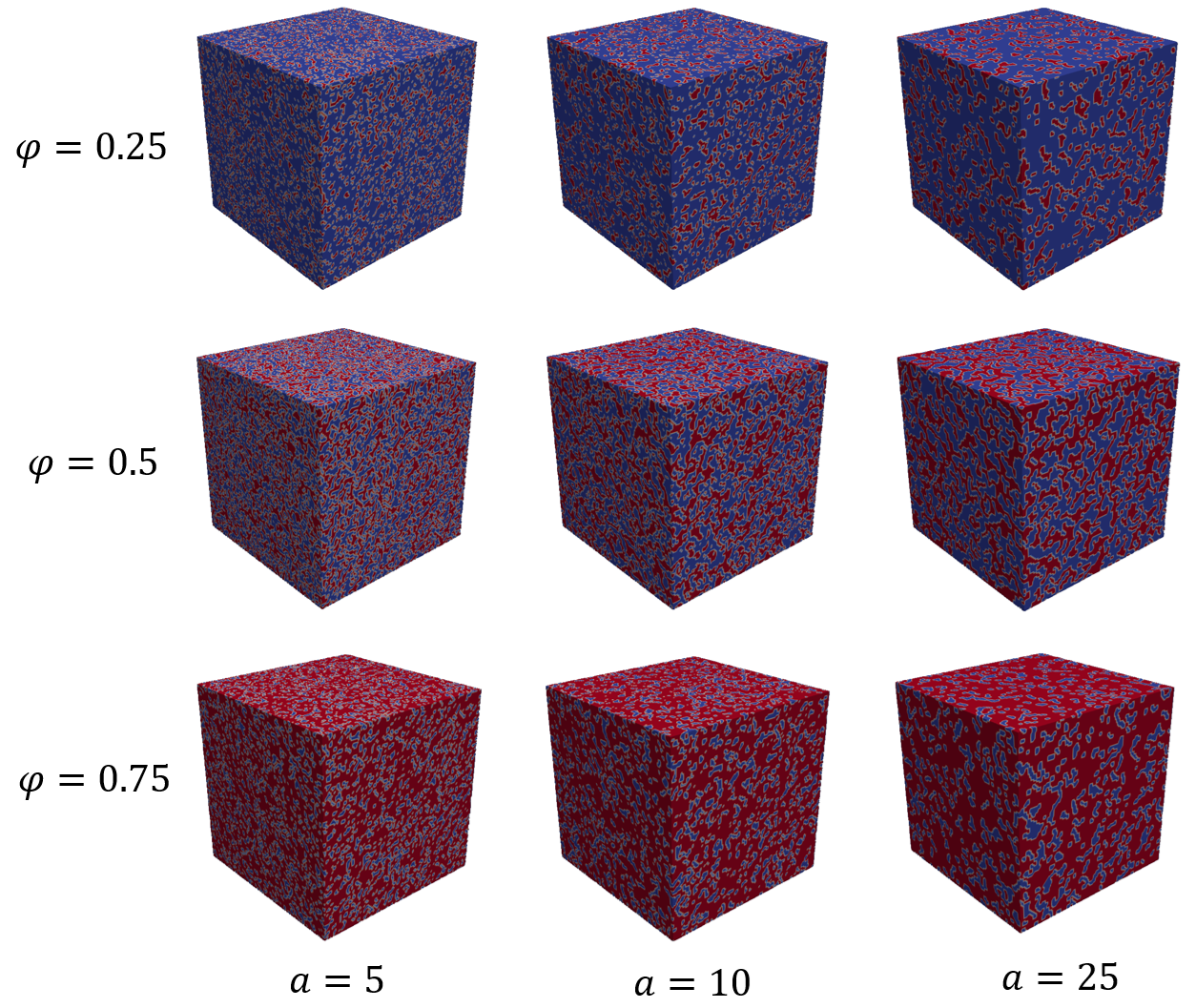}
\end{array}$
\end{center}
\caption{Constructions of the stealthy hyperuniform media with $\phi_1 = 0.25$, 0.5 and 0.75 (top to bottom) and $a = 5$, 10 and 25 voxels, where phase 1 is shown in red color. The linear size of the system is $L = 128$ voxels.} \label{fig_config_stealthy}
\end{figure}

A statistically isotropic stealthy hyperuniform two-phase medium is a subclass of hyperuniform media, which possesses the following spectral density function \cite{torquato2021diffusion}
\begin{equation}
\tilde \chi_{_V}({\bf k}; a) = 0, \quad \text{for}~~ 0\le|{\bf k}|\le K,
\end{equation}
which indicates the scattering in the medium is completely suppressed for a range of wavenumbers up to $K$, and belongs to class-I hyperuniformity. Here, we choose the length scale parameter $a = 2 \pi /K$, which is inversely proportional to the size of the ``exclusion region'' in the Fourier space \cite{shi2023computational}. A smaller $a$ value indicates a larger exclusion region, meaning scattering in the medium is suppressed over a broader range of length scales. We note although stealthy hyperuniform two-phase media have been obtained from stealthy hyperuniform point configurations by placing nonoverlapping spheres at the points \cite{skolnick2023simulated}, 3D realizations of stealthy hyperuniform media directly constructed from target spectral density functions have not been achieved before.   


Figure \ref{fig_chi_stealthy} shows the rescaled autocovariance function $\chi_{_V}({r}; a)/(\phi_1\phi_2)$ (left) and rescaled spectral density function $\tilde \chi_{_V}({k}; a)/(\phi_1\phi_2)$ (right) associated with the stealthy hyperuniform media with varying length scale parameter $a = 5$, 10 and 25 voxels. Increasing $a$ results in a decrease of the exclusion region in $\tilde \chi_{_V}({k}; a)$, which is immediately followed by a peak with decreasing width. The corresponding $\chi_{_V}({r}; a)$ are strongly oscillating functions, indicating significant short- and intermediate-ranged correlations in the system, which are necessary to achieve the complete suppression of scattering over the designated range of wavenumbers.



The generated microstructures for varying $\phi_1$ and $a$ values are shown in Fig. \ref{fig_config_stealthy}. For small $\phi_1$ and large $a$ (i.e., small exclusion region size $K$) (e.g., $\phi = 0.25$, $a = 25$), the phase of interest forms well-defined and uniformly distributed ``ligaments'' with similar sizes. As $\phi_1$ increases while keeping $a$ the same, the particles grow in size and merge into ligaments which eventually form a percolated phase. On the other hand, increasing the exclusion region size $K$ (decreasing $a$) while keeping $\phi$ fixed leads to a finer morphology of the phase with an increasing degree of dispersion. This is because increasing $K$ requires the constructed media to be very ``uniform'' on smaller and smaller scales in real space, which is roughly characterized by parameter $a$, in order to fully suppress the scattering for this range of wavevectors. This requires the realizations to possess finer and more dispersed morphology instead of large clusters. These observations are consistent with previously reported construction results of 2D isotropic stealthy hyperuniform random media \cite{shi2023computational}. Similarly, we observe a high-degree of phase-inversion symmetry in the reconstructed media, consistent with the observation of 2D stealthy hyperuniform media. 





\subsection{Antihyperuniform Two-Phase Media}

\begin{figure}[ht]
\begin{center}
$\begin{array}{c}\\
\includegraphics[width=0.47\textwidth]{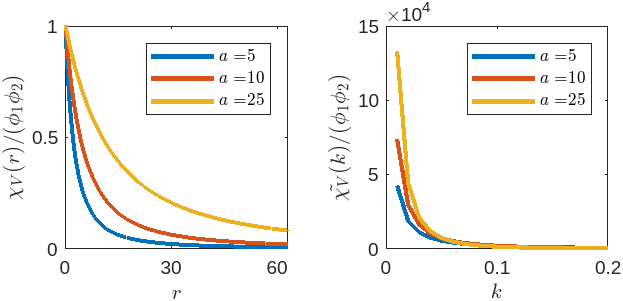}
\end{array}$
\end{center}
\caption{Rescaled autocovariance function $\chi_{_V}({r}; a)/(\phi_1\phi_2)$ (left, c.f. Eq. (\ref{eq_auto_anti})) and rescaled spectral density function $\tilde \chi_{_V}({k}; a)/(\phi_1\phi_2)$ (right, c.f. Eq. (\ref{eq_chi_anti})) associated with the antihyperuniform media with varying length scale parameter $a$. The unit of $r$ is the edge length of a single voxel and the unit of $k$ is $2\pi/L$.} \label{fig_chi_anti}
\end{figure}

\begin{figure}[ht]
\begin{center}
$\begin{array}{c}\\
\includegraphics[width=0.49\textwidth]{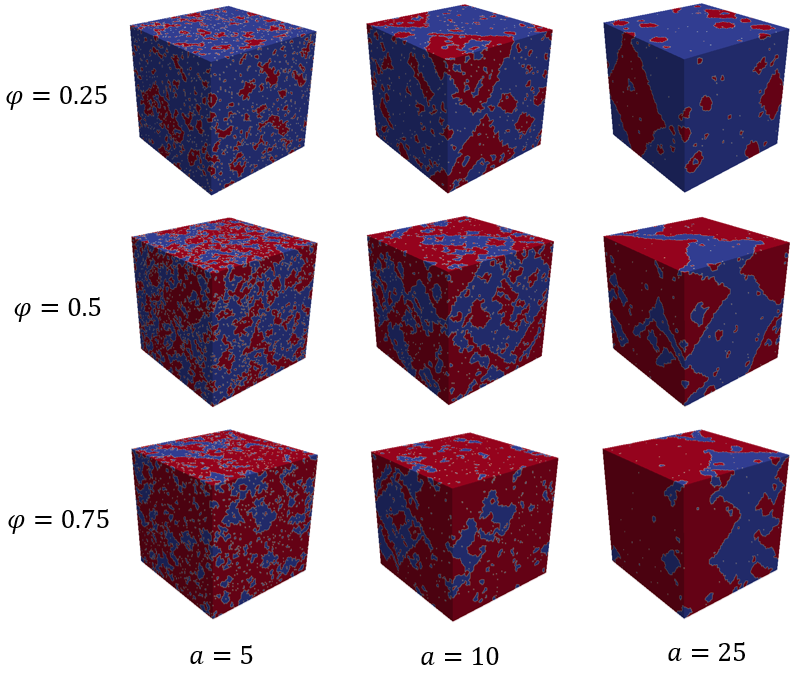}
\end{array}$
\end{center}
\caption{Constructions of the antihyperuniform media with $\phi_1 = 0.25$, 0.5 and 0.75 (top to bottom) and $a = 5$, 10 and 25 voxels, where phase 1 is shown in red color. The linear size of the system is $L = 128$ voxels.} \label{fig_config_anti}
\end{figure}

Torquato \cite{To16b} proposed a model of antihyperuniform two-phase media, which possess a power-law autocovariance function: 
\begin{equation}
\chi_{_V}({r}; a) = \frac{\phi_1\phi_2}{1+2(r/a)+(r/a)^2}.
\label{eq_auto_anti}
\end{equation}
It has been shown \cite{To16b} that the above monotonic functional form meets all of the known necessary realizability conditions on a valid autocovariance function. The associated spectral density function $\tilde \chi_{_V}({k})$ is given by \cite{torquato2021diffusion}
\begin{equation}
\tilde \chi_{_V}({k}; a) = \frac{4\pi a^2}{ka}[Ci(ka)A(ka) + (Si(ka)-\pi/2)B(ka)]
\label{eq_chi_anti}
\end{equation}
where 
\begin{equation}
Ci(x) = \int_0^x dt cos(t)/t, \quad Si(x) = \int_0^x dt cos(t)/t,
\end{equation}
and 
\begin{equation}
A(x) = x \cos(x) + \sin(x), \quad B(x) = x \sin(x) - \cos(x).
\end{equation}
The small-$k$ scaling behavior of $\tilde \chi_{_V}({k})$ can be easily obtained from Eq. (\ref{eq_chi_anti}), i.e., 
\begin{equation}
\tilde \chi_{_V}({k}) \sim 2 \pi^2/k,
\end{equation}
which is consistent with the power-law decay of the autocovariance function $\chi_{_V}({r}) \sim 1/r^2$ for large $r$. Although this model has been employed to analytically investigate the diffusion spreadability of antihyperuniform systems \cite{torquato2021diffusion}, 3D realizations of the antihyperuniform media have never been explicitly constructed.


Figure \ref{fig_chi_anti} shows the rescaled autocovariance function $\chi_{_V}({r}; a)/(\phi_1\phi_2)$ (left), and the rescaled spectral density function $\tilde \chi_{_V}({k}; a)/(\phi_1\phi_2)$ (right) associated with the antihyperuniform media, which diverges as $k\rightarrow 0$ for all different correlation length $a$. Figure \ref{fig_config_anti} shows the constructed two-phase medium. In contrast to the hyperuniform systems, which contain clusters of roughly uniform sizes and spatial distributions, the phase in the constructed antihyperuniform systems forms clusters of dramatically different sizes. For a given $\phi_1$, the largest cluster in the system is determined by the correlation length parameter $a$, which increases in size as $a$ increases. On the other hand, regardless of the $a$ values, all systems contain excess of dispersed voxels (i.e., the smallest possible ``clusters''), which in combination with the rare large clusters, lead to the desired huge density fluctuations and the diverge of zero-wavenumber scattering. Interestingly, the morphology of the phase mimics that of a system at the critical point, i.e., with system-spanning long-lived density fluctuations that strongly suppress scattering. Similarly to previous systems, increasing $\phi_1$ results in enhanced connectivity of the phase of interest. The phase inversion symmetry is observed in the constructions.

\begin{figure*}[ht]
\begin{center}
$\begin{array}{c}\\
\includegraphics[width=0.95\textwidth]{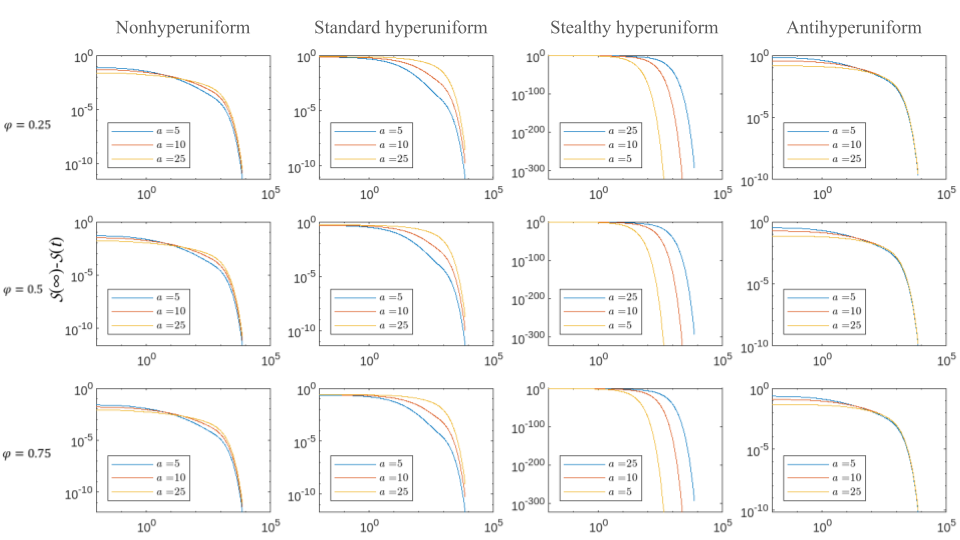}
\end{array}$
\end{center}
\caption{Excess spreadability $\mathcal{S}(t) - \mathcal{S}(\infty)$ for the four classes of constructed two-phase media with varying volume fraction $\phi_1$ and length scale parameter $a$. From left to right: Nonhyperuniform media, standard hyperuniform media, stealthy hyperuniform media, and antihyperuniform media.} \label{fig_spreadability}
\end{figure*}

\begin{table*}[ht]
\centering
\begin{tabular}{c|ccc|ccc|ccc|ccc}
\hline\hline 
 &   & Nonhyper. &   &  & Std. hyper. &  &  & Ste. hyper.& & & Antihyper. &    \\\hline 
$\phi_1$  & $a=5$  & $a=10$ & $a=25$  & $a=5$ & $a=10$ & $a=25$ & $a=5$ & $a=10$ & $a=25$ & $a=5$ & $a=10$ & $a=25$     \\\hline 
0.25  & 0.9309  & 0.9087 & 0.8631  & 0.4378 & 0.4261 & 0.4014 & 0.2796 & 0.2758 & 0.2403 & 1.679 & 1.1283 & 0.9467 \\\hline 
0.5   & 0.6625 & 0.6431 & 0.6308  & 0.9603 & 0.09539 & 0.9218 & 0.03793 & 0.03702 & 0.03538 & 0.8625 & 0.7736 & 0.6622\\\hline 
\hline
\end{tabular}
\caption{Dimensionless fluid permeability $k/a^2$ of the constructed two-phase media with varying phase volume fraction $\phi_1$ and length scale parameter $a$, estimated using Eq. (\ref{eq_k_est}), where the permeability $k_0$ of a simple-cubic packing of hard spheres has been used as the reference \cite{jung2005fluid}.}
\label{tab_k}
\end{table*}

\section{Transport Properties of Constructed Media}

In this section, we obtain the transport properties including the time-dependent diffusion spreadability $\mathcal{S}(t)$ and fluid permeability $k$ of the constructed two-phase media via their spectral density function ${\tilde \chi}_{_V}({k})$. Specifically, $\mathcal{S}(t)$ is computed rigorously using Eq. (\ref{eq_St}) and $k$ is estimated using Eq. (\ref{eq_k_est}) and $k_0$ for the simple cubic packing of congruent hard spheres \cite{jung2005fluid}. We note since both the diffusion spreadability and fluid permeability are directly obtained from the parameterized function ${\tilde \chi}_{_V}({k}; a)$, these quantities also explicitly depend on the length scale parameter $a$. This offers an efficient computational framework that allows one to obtain targeted transport properties $\mathcal{S}(t; a)$ and $k(a)$ by optimizing parameter $a$ for a given class of two-phase media. 

Figure \ref{fig_spreadability} shows the results of excess spreadability $\mathcal{S}(t) - \mathcal{S}(\infty)$ for the four classes of constructed two-phase media, including nonhyperuniform (Debye random media), standard disordered hyperuniform, stealthy hyperuniform and antihyperuniform materials, with varying volume fraction $\phi_1$ and length scale parameter $a$. Torquato \cite{torquato2021diffusion} showed that the long-time excess spreadability for the above media possesses the following scaling behavior: $\sim t^{-3/2}$ for Debye random media, $\sim t^{-5/2}$ for standard hyperuniform media, $\sim e^{-\theta t}/t$ for stealthy hyperuniform media, and $\sim t^{-1}$ for antihyperuniform media, which are in complete agreement with our numerical results. Importantly, our results indicate that varying the length-scale parameter $a$ within each class of functions can also lead to orders of magnitude variation of ${\cal S}(t)$ at intermediate and long time scales. This indicates the feasibility of employing the parameterized autocovariance functions (or equivalently spectral density functions) as effective reduced-dimension representations for the full microstructure space for time-dependent diffusive transport property engineering. 

Table \ref{tab_k} reports the dimensionless fluid permeability $k/a^2$ of the constructed two-phase media with varying phase volume fraction $\phi_1$ and length scale parameter $a$, estimated using Eq. (\ref{eq_k_est}), where the permeability $k_0$ of a simple-cubic packing of hard spheres with the corresponding $\phi_1$ has been used as the reference \cite{jung2005fluid}. Here we consider phase 1 (red) is the solid phase and phase 2 (blue) is the pore phase, and only report the results of $k$ for $\phi_1 = 0.25$ and 0.5, for which the pore phase is guaranteed to percolate to support a non-zero permeability. It can be seen that for the same class of media, increasing solid volume fraction $\phi_1$ (i.e., decreasing the porosity) leads to a decrease of permeability, as expected. On the other hand, increasing the length scale parameter $a$ also results in a decrease of the dimensionless permeability $k/a^2$ (although the actual permeability $k$ increases with increasing $a$). In general, permeability sensitively depends on many microstructural factors such as interracial area, pore space connectivity, spatial distribution of bottlenecks, etc.  Increasing $a$ generally leads to a coarsen morphology with more connected pore space and less interracial area, which give rise to the increasing $k$. However, such increase of $k$ is not as fast as $a^2$ in the disordered media considered here. Interestingly, we find that the antihyperuniform media provide the largest dimensionless $k/a^2$ while the stealthy hyperuniform media possess the smallest $k/a^2$. This is in contrast to the spreadability results, for which among the four classes of two-phase media, the stealthy hyperuniform and antihyperuniform ones respectively achieve the most and least efficient diffusive transport.    








\section{Conclusions and Discussion}




In this work, we presented an efficient Fourier-space based computational framework to construct statistically isotropic two-phase random media in $\mathbb{R}^3$ based on prescribed target spectral density functions ${\tilde \chi}_{_V}({k})$. We have employed a variety of analytical models of ${\tilde \chi}_{_V}({k})$ which satisfy all known necessary conditions to construct representative nonhyperuniform, standard hyperuniform, stealthy hyperuniform, and antihyperuniform two-phase random media at various phase volume fractions. Within each family of two-phase media, the corresponding spectral density function (or equivalently the autocovariance function) contains a length scale parameter $a$, which allows us to tune the correlations in the system across length scales via the targeted functions to generate a rich spectrum of distinct microstructures. While the analytical model functions $\tilde \chi_{_V}({k}; a)$ we considered here have been employed to model transport properties of two-phase random media \cite{torquato2021diffusion}, the explicit 3D realizations of the majority of these functions such as the antihyperuniform media have never been obtained before. We found that the antihyperuniform media constructed here typically contain a mixture of a single large, irregular cluster with many much smaller clusters of different morphologies, mimicking the structure of a system at the critical point. In a future study, we will undertake a comprehensive study of the clustering and percolation properties of all of our constructed media.





The spectral density function ${\tilde \chi}_{_V}({k})$ also enables accurate determination of various transport properties such as the time-dependent diffusion spreadability ${\cal S}(t)$, and estimates of the fluid permeability $k$ of the constructed media. In Ref. \cite{torquato2021diffusion}, Torquato has shown that the long-time asymptotic scaling behavior of ${\cal S}(t)$ only depends on the functional form of ${\tilde \chi}_{_V}({k})$, with the stealthy hyperuniform and antihyperuniform media respectively achieving the most and least efficient diffusive transport. Our numerical results here further confirmed the theoretical predictions in Ref. \cite{torquato2021diffusion} that varying the length-scale parameter within each class of ${\tilde \chi}_{_V}({k})$ functions can also lead to orders of magnitude variation of ${\cal S}(t)$ at intermediate and long time scales in disordered hyperuniform, nonhyperuniform and antihyperuniform media. For the fluid permeability, we found that increasing solid volume fraction $\phi_1$ and correlation length $a$ in the constructed media can lead to a decrease in the dimensionless fluid permeability $k/a^2$. Interestingly, the antihyperuniform media possess the largest $k/a^2$ among the four classes of materials with the same $\phi_1$ and $a$, while the stealthy hyperuniform media possess the smallest $k/a^2$. We note this trend is consistent with previous reported sharp bounds of $k/a^2$ \cite{torquato2020predicting} and a recent comprehensive study of triply periodic bicontinuous materials \cite{torquato2024microstructural}. Since both the diffusion spreadability and fluid permeability are directly obtained from the parameterized function ${\tilde \chi}_{_V}({k}; a)$, these quantities also explicitly depend on the length scale parameter $a$. This offers an efficient computational framework that allows one to obtain targeted transport properties $\mathcal{S}(t; a)$ and $k(a)$ by optimizing parameter $a$ for a given class of two-phase media. 

\begin{figure*}[ht]
\begin{center}
$\begin{array}{c}\\
\includegraphics[width=0.6\textwidth]{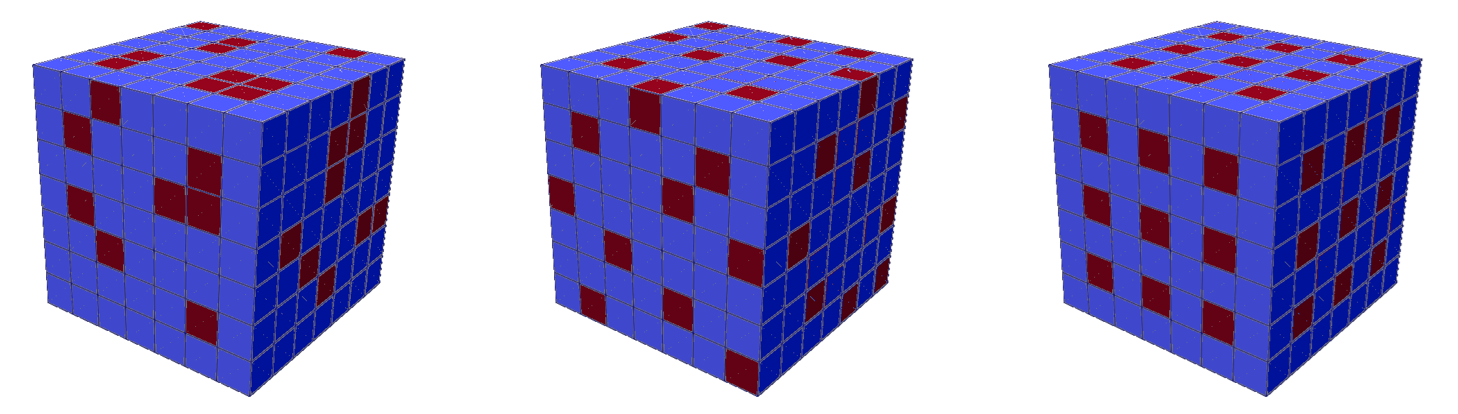}
\end{array}$
\end{center}
\caption{Constructed microstructures with $\phi = 1/8$ associated with increasing $\chi$. From left to right, $\chi = 0.05$, 0.2 and 0.4, respectively. The linear size of the system is $L = 8$ voxels. The smaller system size allows fast convergence of the construction algorithm.} \label{fig_large_circle}
\end{figure*}

Our current results indicate the feasibility of employing parameterized ${\tilde \chi}_{_V}({k})$ for designing composites with targeted transport properties, which can be readily adapted to design two-phase media with desirable dynamic wave properties. This can be achieved by leveraging recently developed predictive formulations including nonlocal theories for the effective dynamic elastic moduli and dielectric constant \cite{kim2020multifunctional, kim2020effective, torquato2021nonlocal}. Such theories rigorously connect effective properties of the media to their spectral density ${\tilde \chi}_{_V}({\bf k})$, allowing us to achieve desirable material properties by tuning and optimizing a targeted parameterized spectral density. The parameterized ${\tilde \chi}_{_V}({k})$ can include both the correlation length as a tuning parameter as in the current work, or a combination of a number of realizable basis functions that allows engineering of various microstructural features \cite{jiao2007modeling}. The associated microstructure can then be obtained using the Fourier-space construction procedure. In future work, we will explore this framework to design disordered hyperuniform composites with targeted electromagnetic wave characteristics.

\begin{acknowledgments}
This work was supported by the Army Research Office under Cooperative Agreement Number W911NF-22-2-0103.
\end{acknowledgments}
\smallskip

\appendix
\section{Effects of Increasing $K$ for Stealthy Hyperuniform Media}


In the case of stealthy hyperuniform media, we have shown that increasing the size of the exclusion region $K$ results in finer morphologies and enhances dispersion of individual pixels in the constructed microstructures. As discussed in Sec. IV.C, increasing $K$ requires suppression of local volume fraction fluctuations on smaller length scales. Here, we further show that increasing $K$ (or equivalently $\chi$) can lead highly ordered configurations of the voxels, as if the voxels are hard ``particles'' on a lattice, which was also observed in 2D media \cite{shi2023computational}. 


In particular, we investigate the evolution of microstructures as the exclusion regions with increasing size $K$ on the cubic lattice. Without loss of generality, we consider a system with $\phi = 1/8$, $L = 8$ voxels and $N = 64$, which possesses an ordered microstructure with the individual ``red'' voxels arranged on a perfect cubic lattice with a lattice constant $a = 2$ voxels. The smaller system size allows fast convergence of the construction algorithm. Figure \ref{fig_large_circle} shows the constructed microstructures associated with increasing $K$, and thus increasing $\chi$. It can be seen that for low $\chi$ values, the system is in the ``random-medium'' regime, and the resulting microstructures contain clusters of red voxels. As $\chi$ increases, the red voxels behave more like ``particles'' with an increasing degree of ``repulsion'', and the resulting microstructures contain distributions of individual red voxels with increasing local order. An almost perfect cubic-lattice configuration of the red voxels is obtained at $\chi = 0.4$. These results are consistent with 2D stealthy hyperuniform media with increasing $\chi$ values \cite{shi2023computational}.
\bigskip


\bibliography{reference}

\end{document}